# Surface Fluctuations of an Aging Colloidal Suspension: Evidence for Intermittent Quakes.

Alexandre Mamane, Christian Frétigny, François Lequeux, and Laurence Talini,.

Laboratoire PPMD, Université Pierre et Marie Curie-Paris6, UMR CNRS 7615, ESPCI – 10 rue Vauquelin, 75005 Paris, France.

Abstract:

We present measurements of the thermal fluctuations of the free surface of an aging colloidal suspension, Laponite. The technique consists in measuring the fluctuations of the position of a laser beam that reflects from the free surface. Analysing the data statistics, we show that, as the fluid ages, the dynamics becomes intermittent. The intermittent events correspond to large changes in the local slope of the free surface over a few milliseconds. We show that those quakes are uncorrelated, although they are kept in memory by the surface over short time scales.

Heterogeneous dynamics is encountered in very different situations such as turbulent flows of liquids [1], granular flows [2], chemical reactions [3], aging of systems that are close to the glass transition [4]… The heterogeneities are spatial as well as temporal. For the latter aspect, intermittent behaviour consist in non uniform microscopic fluctuations over time: periods of relative calm are interspersed with periods of intense activity.

In the case of systems close to the glass transition, dynamical heterogeneities have been evidenced both numerically and experimentally in the past years [5], despite they are not predicted by most theoretical models for the glass transition. The co-existence of slow and fast regions has thus been observed in glass formers of different natures [6]. Intermittencies have been documented in systems such as polymer films at a temperature slightly above the glass transition temperature [7], a polymer glass quenched below the glass transition temperature [8], or glassy phases formed from colloidal suspensions with a net attractive interaction between particles [9-10]. However, experimental evidence for heterogeneous dynamics remains difficult since the system must be probed at a small enough length scale and without averaging the measurements over time. Owing to the scarcity of quantitative measurements of intermittent behavior, it is not possible to have a clear overview of the length and time scales it is associated with, and further experimental documentation on intermittencies is needed to elucidate the underlying physical mechanisms.

Glassy phases of colloidal suspensions provide easier-to-study systems since the involved length scales are larger than in the ones in polymeric systems. Dynamical heterogeneities in those systems have been evidenced using either measurements of tracer diffusion [11], or light scattering techniques [9;10;12]. In the former case, the main drawback is that the technique is invasive, whereas in the latter cases, the measurements are performed over large volumes (a few mm$^3$), and are therefore suited only for systems in which the length scale of the heterogeneities is large. Furthermore, the data treatment being far from straightforward, it is difficult to identify the changes induced by the events at a microscopic scale, and thus to understand the nature of the rearrangements. Herein, we present measurements of intermittencies on the free surface of a colloidal glassy phase at a lengthscale much larger than the microscopic one. We show that dynamical heterogeneities during the aging of the glassy phase can be evidenced using an analysis of the thermal fluctuations of the free surface of the suspension. Intermittencies are signalled by non-Gaussian fluctuations of the free surface. We report measurements of the distribution of the time intervals between the corresponding events, and show how the surface is affected by each event, using a simple analysis of the raw signal.

The glassy material is an aqueous suspension of a synthetic clay, Laponite (RD, Rockwood). The suspensions are prepared by mixing 2.5% w/w Laponite powder with water added with hydroxide sodium (concentration $10^{-4}$M). In those conditions, Laponite is known to form an isotropic phase that is out of equilibrium and ages. When fully hydrated, the Laponite platelets are disc-like, of diameter 25nm and height 1nm. The anisotropy of the particles as well as the charges they bear (that are negative on the faces and positive on the edges of the platelets in our experimental conditions) leads to complex particle interactions, and the nature of the formed phase at so small a volume fraction is still a matter for debate. Some authors describe it as a gel with a house of cards structure [13], whereas others invoke a repulsive net interaction between particles, and thus the formation of a colloidal glass [14]. Whatever its exact nature, the formed phase is out of equilibrium, and, when at rest, its relaxation time increases with time [15]. In the presented experiments, special care has been taken to ensure that the aging results from a physical process and not from a slow hydration of the clay platelets [16]. Once prepared, the suspension was kept one month before use: for two weeks it was mixed for about 1 hour every two days using a magnetic stirrer, it was then kept at rest for two more weeks. Within those conditions, the Laponite platelets are fully hydrated; we have in particular checked that the state of the suspension only depends on the time elapsed after it has been rejuvenated, i.e. strongly stirred for at least half an hour. In the following we denote as $t_w$ the aging time (or waiting time) whose origin coincides with the rejuvenation, whereas the origin of time $t$ is the beginning of the measurement. The age of the fluid during a measurement run is therefore $t_w + t$. Since the duration of one run is always small compared to the aging time, we will consider in the following that the age during a measurement is simply given by $t_w$.

Once rejuvenated, the Laponite suspension is poured in a cell (diameter 5cm), the total suspension height being smaller than 2mm. The sample is covered to prevent water evaporation. The surface fluctuations of the suspension have been measured at different waiting times (from 1 to 35 hours) using the technique of Surface Fluctuation Specular Reflection (SFSR) [17]. It consists in measuring the fluctuations in the position of a laser beam that is reflected from the free surface of the suspension at rest. At the first order, and as schematized in Fig. 1, the position of the reflected beam depends on the local slope of the surface. More precisely, SFSR consists in measuring the average slope of the free surface over the surface of the laser beam, i.e. over a circular spot, which is of diameter 60µm in the present work. The fluctuations of that position can therefore be related to the propagation of the thermally excited waves at the probed surface. The spatial modes that significantly contribute to the signal are the ones of wavelengths larger than the beam size, which is 60µm at the beam waist. Figure 1 briefly describes the experimental set-up: the beam is focused on the free surface and its reflected part is centered on the two quadrants of a photodiode; the quadrants are connected to a current to voltage converter and we measure the signal $S(t)$, which is the difference between the voltages delivered by the two quadrants, after amplification and anti-aliasing filtering. The sensitivity is such that a signal of 1mV corresponds to a variation of $3 \times 10^{-8}$ rad in the slope of the free surface. In the presented experiments, $S(t)$ is digitized at a sampling rate of 2kHz; the duration of one run is 512s, more than $10^6$ data points are thus acquired during one run. The experimental set-up is fully detailed elsewhere [17].

We have shown in a previous work how, for systems that are in thermodynamical equilibrium, the power spectrum density of $S(t)$ could be related to the power spectrum density of the surface height, and thus to the properties of the probed medium, through a fluctuation dissipation relation. Such a relation does not a priori hold in the case of Laponite suspensions, and we have instead analysed the raw signal $S(t)$ following a process specially developed to evidence dynamical heterogeneities. We have nevertheless controlled as well the variations of the power spectrum density of $S(t)$. As will be detailed in a forthcoming paper, the power spectrum density of Laponite is characteristic of soft gel-like materials, on the surface of which propagate, according to the frequency, either elastic-like or capillary-like thermal waves [18]. The variations with the aging time of the power spectrum densities show in particular an increase of the elastic modulus of Laponite, as reported in other works [19]; those variations, reflecting the elastic nature of the sample, moreover evidence that, over the experimental time scales, no significant modification of the Laponite/air interface occurs by syneretic or sedimentation processes.

The intermittencies consisting in rare events, a sensitive statistical estimator is needed in order to detect them [20]. We have used the convolution of $S(t)$ and the function $M(t,\tau)$, $\Delta(t,\tau) = S(t) * M(t,\tau)$. Function $M(t,\tau)$ is the Mexican hat wavelet, i.e. the normalized, second derivative of a Gaussian function, of variance $\tau^2/3$. For a given time $\tau$, the estimator $\Delta(t,\tau)$ provides a measurement of the second order difference of the signal at timescale $\tau$: large values of $\Delta(t,\tau)$ correspond to bursts of activity at the free surface over times of the order of $\tau$. We have found that non-Gaussian distributions of $\Delta$ appear for $\tau$ smaller than a few milliseconds. In the following, the time $\tau$ is kept at a constant value of 2ms, which is large enough compared to the sampling time. We have checked that, for the same value of $\tau$, function $\Delta$ for a viscous fluid (silicon oil) presents a Gaussian distribution. Experiments performed with a sample of Laponite that had been kept at rest for only 15 days, and was therefore probably not fully hydrated, also showed a Gaussian behavior after the same data treatment. Special care has also been taken to ensure that small deviations of the laser centering on the two quadrants of the photodiode do not induce significant modifications in the distribution of $\Delta$.

Figure 2 shows the distribution of function $\Delta(t, \tau = 2ms)$ at different ages of the Laponite suspension. For a "young" Laponite suspension of a few hours, the distributions of $\Delta(t)$ are well described by a Gaussian curve, whereas they present a systematic deviation from Gaussian curves at larger $t_w$. At those times, the large tails observed in the distributions indicate that bursts of activity occur over time scales close to $\tau$, more frequently than they would if the distributions were Gaussian. Note that the growing flatness of the distribution of $\Delta$ is associated with a decrease of its width with the aging time; as shown in the inset of Fig. 2, the standard deviation of the distribution of $\Delta$ first strongly decreases with the aging time and then seems to reach a plateau. Therefore, in the first 25 hours, the surface fluctuations get smaller as Laponite ages, reflecting the overall slowing down of the dynamics; however we also observe the occurrence of fluctuations larger than the average. This is different from what has been observed by particle tracking in a clay suspension [11]. The suspension was an aqueous hectorite suspension and the diffusive motion of micron sized tracers was measured. Contrary to the present results, as the fluid ages, the distribution of the tracer mean square displacements becomes non-Gaussian owing to the increase in the proportion of smaller displacements. That behavior was associated with the increasing caging of the particles. Laponite is as well a hectorite clay (although synthetic and of much smaller platelet size), direct comparison of the experimental data is however difficult, owing to the differences in the nature of the fluctuations probed. Furthermore, it has been recently shown that the sensitivity to dynamical heterogeneities in glassy colloidal suspensions strongly depends on the tracer size [21].

We have further characterized the events causing the deviation from a Gaussian distribution of the function $\Delta$. The occurrence of each of the abnormally frequent bursts can be detected by reporting the times for which $\Delta$ is larger than a threshold value. For each value of $t_w$, the threshold value has been chosen such that the Gaussian fit of the distribution yields 1 for that value. Figure 3 shows an example of events, identified in the raw signal $S(t)$, and corresponding to values of $\Delta$ larger than the threshold. They consist in large amplitude events, and thus to large variations of the slope of the free surface occurring within a few milliseconds. For $t_w \geq 14$h, between 1000 and 2500 events are detected in the signal of duration 512s, without any significant variation in their number with the aging time. The distribution of the time intervals between two events is shown in the inset of Figure 3. We observe Poissonian statistics, with a mean characteristic time that first decreases with the aging time, and further reaches a value close to 0.3s for $t_w \geq 14$h. Power law distributions of the time intervals between dynamic heterogeneities have been reported in other systems close to the glass transition: a polymer glass [7], and a coagulating suspension [9]. Such statistics prove the existence of correlations between events, which is not the case in our system. Indeed, the length scale at which the system is probed is expected to greatly influence the observed statistics. It has for instance been shown that, in the case of a system constituted by $N$ subsystems, each being described in the frame of a model of traps, either exponential, stretched exponential or power law distributions of the time intervals between events could be predicted, according to the number of domains probed [22]. We furthermore show in the following that, although the events do not seem to be correlated, short time correlations can however be observed in our system.

We have analysed the modifications on the free surface induced by the events. For each event detected in the raw signal, we have computed the difference in the signal averaged over a time $T$ before and after the event. To ensure the large amplitudes induced by one event are not taken into account, the time lapse over which the signal is averaged before the event ends 1 ms before beginning of the event (defined as the data point where $\Delta$ is larger than the threshold value); in a symmetrical way, the time lapse after an event starts 1 ms after the end of the event, signalled by the last value of $\Delta$ larger than the threshold. Examples are shown in Fig. 3, where, for $T = 10$ms, the intervals over which we average after and before an event appear as horizontal segments. Events separated by a time interval smaller than $T$ are not taken into account in the computation. We thus probe the difference in the slope of the free surface. In Figure 4 are displayed the distributions of the signal difference averaged over time $T$. Figure 4a shows the distributions at a given time $T = 10$ms, and for different aging times, whereas Figure 4b shows the distributions at a given aging time ($t_w = 16$h) and for different times $T$. For small aging times ($t_w < 14$h), the distribution is Gaussian, showing that the slope of the free surface before an event is not correlated to the one after the same event. Since only a few events (less than 200 per run) are detected for those ages, the statistics is poor. When a large number of intermittencies occur ($t_w \geq 14h$), we observe a change in the shape of the distribution. As in the example of Fig. 4a for $t_w$ =16h, the width of the distribution increases and it exhibit two symmetric bumps. Those bumps demonstrate that the occurrence of an event favors a strong change in the slope of the free surface. An event therefore results in a strong modification of the free surface that is kept in memory for a finite time; the events may be thought of as quakes, similarly to the "cracks" described in an aging system of different nature [9]. Indeed, the smaller the time $T$ the more markedly do the bumps appear, whereas, as $T$ is increased, the bumps fade away and the distribution becomes Gaussian, showing the memory of a quake disappears for times larger than 20 ms (Fig. 4b).

Figure 4b also shows that the distributions of the signal differences further evolve with the aging time. As the fluid ages, the bumps fade away, while the width of the distribution decreases. The distribution actually becomes Gaussian again (Fig. 4b), although the number of intermittencies does not significantly vary. At those aging times, the time scale of the correlation between the fluctuations before and after and event might be smaller than the resolution of the measurement. Shorter correlation times for larger aging times may seem counterintuitive, however, the elastic properties are doubtlessly involved in the quake process: the increase of the elastic modulus with the aging time is likely to change the time over which an event affects the surface.

In summary, we present experimental evidence for intermittencies during the aging of a glassy colloidal suspension. We have used the reflection of a laser beam on the free surface of the suspension, the position of which is sensitive, at the first order, to the local slope of the surface. We show that, at large enough aging times, the thermally induced fluctuations of the surface exhibit bursts of activity, characterized by non-Gaussian dynamics. We have found that the corresponding event induce large changes in the local slope of the free surface – at a scale larger than the 60μm of the beam size - and are uncorrelated. The surface has nevertheless a short-time memory of each of those quakes. The involved timescales are of the order of a few milliseconds at the spatial scale we probe. Investigation of the spatial aspect of the dynamical heterogeneities is possible with SFSR, probing smaller length scales as well as investigating the spatial correlations are parts of our future works. Finally, comparison of our results with ones obtained in the bulk of a colloidal glassy suspension could provide information on the influence of the vicinity of a surface, close to which faster aging dynamics is expected [23].

We thank L. Berthier and H. Montes for fruitful discussions. L.T gratefully acknowledges the support of CNRS for her full time research year.

Figure captions

Figure 1: Schematized view of the experimental set-up. A laser beam is focused onto the free surface of the Laponite sample, its reflected part is centered on the two quadrants of a photodiode. We use the reflections of the beam on two sides of a prism. At the first order, the position of the reflected beam is sensitive to the slope of the free surface.

Figure 2: Histograms of the values of the function $\Delta(t,\tau)$ that gives a measurement of the second order difference of the signal at timescale $\tau$. Time $\tau$ is set at 2ms and the full lines with different gray scales correspond to different aging times (the larger the aging time, the lighter the line): $t_w$ =1h, $t_w$ = 10h and $t_w$ =30h. The Gaussian fit to the distribution at $t_w$ =1h is also shown (black line). The values of $\Delta$ have been normalized by the standard deviation of the distribution of $t_w$. In the inset are shown the variations of that standard deviation, which has been normalized by the value at $t_w$ =1h.

Figure 3: Raw signal $S(t)$ for $t_w$ =30h. In order to show the intermittent events, only a small part of the run is shown (about 100ms out of 512s). The two large amplitude events correspond to values of $\Delta$ larger than a threshold value, which has been chosen such that the Gaussian fit of the distribution of $\Delta$ yields 1. The horizontal segments indicate, for each event, the time lapses over which the signal is averaged to compute the difference of the signal before and after an event. The inset shows the distribution of the time intervals between the events for $t_w \geq 14h$. The experimental data is well described by an exponential function (full line).

Figure 4: Distribution of the differences of the signal before and after each event (as indicated in Fig. 3) averaged over a time $T$=10ms and for different aging times (a) and at a given aging time $t_w$ =16h and for different durations $T$ (b). The different gray lines correspond to different aging times in (a) (from darker to lighter gray: $t_w$ =10h, 16h and 30h) and times $T$ in (b) (from darker to lighter gray: $T$= 5ms and 20ms).

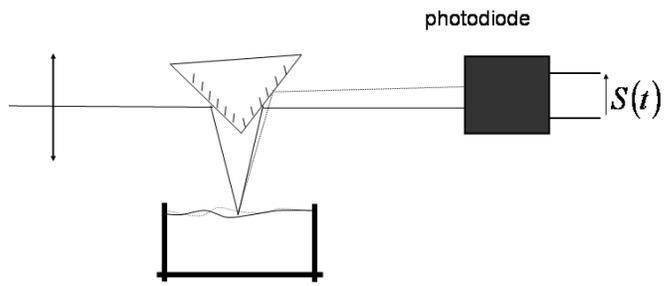

Figure 1

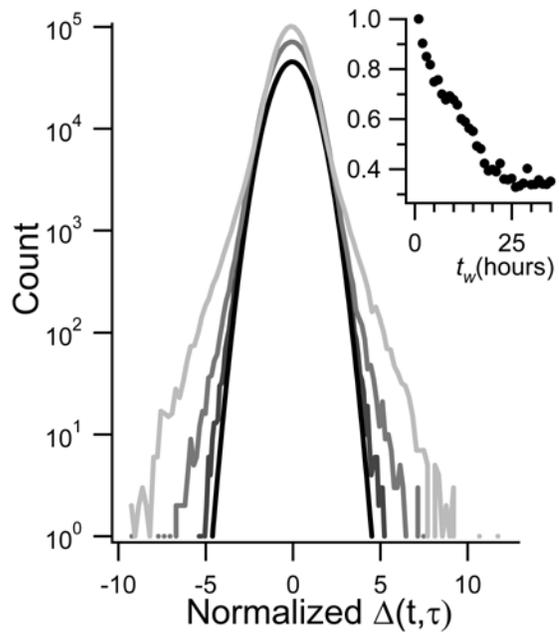

Figure 2

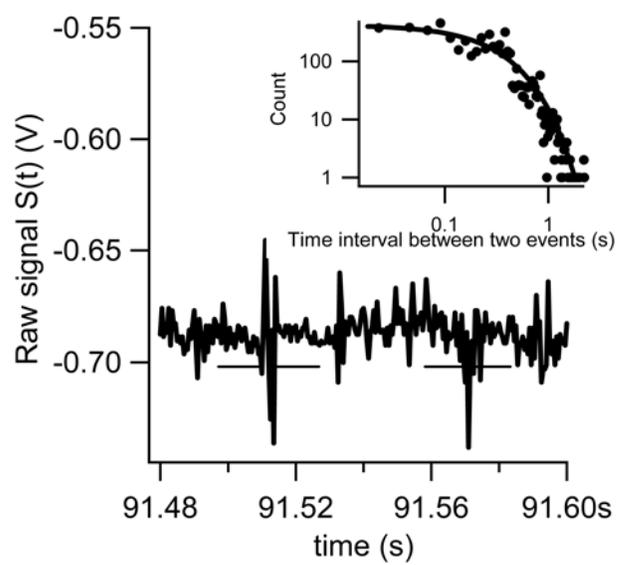

Figure 3

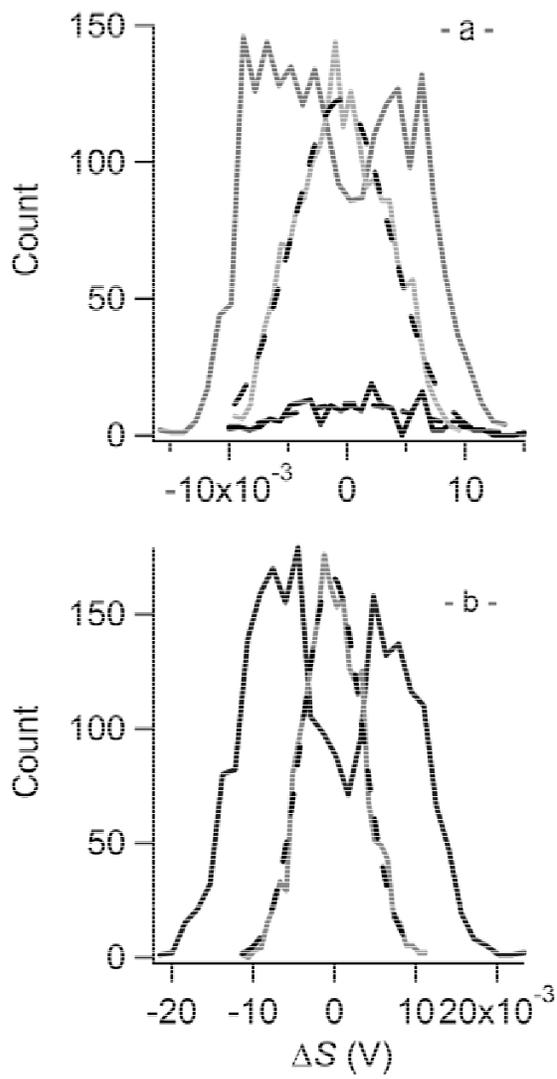

Figure 4